\documentclass[12pt]{article}
\usepackage{epsfig}
\usepackage{citesort}
\usepackage{amssymb,amsfonts}
\newcommand{\beq}{\begin{equation}}
\newcommand{\eeq}{\end{equation}}
\newcommand{\bea}{\begin{eqnarray}}
\newcommand{\eea}{\end{eqnarray}}

\begin{document}
\title{\bf Phases of QCD: lattice thermodynamics and a field theoretical 
model\footnote{Work supported in part by BMBF and INFN}}
\author{Claudia Ratti$^{a,b}$, Michael A. Thaler$^a$ and Wolfram Weise$^a$ 
\\\small{$^a$Physik-Department, Technische Universit\"at M\"unchen, D-85747 
Garching, Germany}\\ 
\small{$^b$ECT$^*$, I-38050 Villazzano (Trento), Italy}\\  
{\small\it (Email: Claudia\_Ratti@ph.tum.de, Michael\_Thaler@ph.tum.de, 
Wolfram\_Weise@ph.tum.de)}}
\date{\today}
\maketitle
\begin{abstract}
We investigate three-colour QCD thermodynamics at finite quark chemical 
potential. Lattice QCD results are compared with a generalized Nambu
Jona-Lasinio model in which quarks couple simultaneously to the chiral 
condensate and to a background temporal gauge field representing Polyakov
loop dynamics. This so-called PNJL model thus includes features of both 
deconfinement and 
chiral symmetry restoration. The parameters of the Polyakov loop effective
potential 
are fixed in the pure gauge sector.
The chiral condensate and the Polyakov loop as functions
of temperature and quark chemical potential are calculated by minimizing the
thermodynamic potential of the system. The resulting equation of state, 
(scaled) pressure difference and quark number density at finite quark chemical potential are then
confronted with corresponding Lattice QCD data.
\end{abstract}
\vskip 2cm
PACS: 12.38.Aw, 12.38.Mh
\vfill\eject

\section{Introduction}
Recent years have seen an expansion of activities devoted to the study of the QCD
phase diagram. Heavy-ion experiments are looking for signals of the 
Quark-Gluon Plasma. Large-scale 
lattice simulations at finite temperature have become a principal tool for 
investigating the pattern of phases in QCD. Accurate computations of lattice 
QCD thermodynamics in the pure gauge sector have been performed. First 
results at finite quark chemical 
potential are available. The equation of state of strongly interacting
matter is now at hand as a function of temperature $T$ and in a limited range 
of quark
chemical potential $\mu$. Improved multi-parameter 
re-weighting techniques~\cite{Fodor:2002km,Fodor:2001pe}, Taylor 
series expansion methods~\cite{Allton:2002zi,Allton:2003vx,Allton:2005gk} and
analytic continuation from imaginary chemical 
potential~\cite{Laermann1,deForcrand:2003hx,delia1,delia2} provide
lattice data for the pressure, entropy density, quark density and selected
susceptibilities.

A straightforward interpretation of these data in terms of QCD 
perturbation theory does not work because of poor convergence at any 
temperature of practical interest~\cite{Arnold:1994eb,Zhai:1995ac}. In order
to overcome this problem, resummation schemes have been proposed, based for
example on the Hard Thermal Loop (HTL) approach
\cite{Braaten:1991gm,Frenkel:1991ts,Blaizot:1993be,Andersen:1999fw,Andersen:2000zn,Andersen:2002ey,Blaizot:1999ip,Blaizot:1999ap,Blaizot:2000fc} or on 
dimensionally reduced screened perturbation theory (DRSPT)
\cite{Kajantie:2002wa,Blaizot:2003iq,Ipp:2003yz}. However, these approaches 
still
give reliable results only for temperatures $T\gtrsim 2.5\,T_c$, far above the
critical temperature $T_c\sim$ 0.2 GeV.
At these high temperatures, the HTL approach motivates and justifies a picture of 
weakly interacting quasiparticles, as determined by the HTL propagators.

In order to extend such descriptions to lower temperatures closer to $T_c$, 
various 
models have been proposed. 
Early attempts were based on the MIT bag model
\cite{Engels:1982ei}. More sophisticated approaches
became necessary when more precise lattice data appeared. Various aspects of QCD thermodynamics have been investigated in
terms of quasiparticle models based on perturbative calculations carried out
in the HTL scheme
\cite{Peshier:1995ty,Levai:1997yx,Peshier:1999ww,Szabo:2003kg,Bluhm:2004xn,Bluhm:2004xn2}, in
terms of a condensate of
$Z_3$ Wilson lines~\cite{Pisarski:2000eq}, by refined quasiparticle
models based on the HTL-resummed entropy and extensions 
thereof~\cite{Rebhan:2003wn}, by an improved version with a temperature-dependent number of active degrees of freedom~\cite{Schneider:2001nf,Thaler:2003uz}, by an evaporation model
of the gluon condensate~\cite{Drago:2001gd}, by quasiparticle models
formulated in dynamical terms~\cite{Ivanov:2004gq}, and by hadron resonance
gas models below the critical temperature~\cite{Karsch:2003zq} 
(for a recent review see~\cite{Rischke:2003mt}). 
 
In this paper, we study the thermodynamics of two-flavour QCD at finite 
quark chemical potential. Our investigation is based on a synthesis of a Nambu 
Jona-Lasinio (NJL) model
\cite{Nambu:1961tp,Nambu:1961fr,Vogl:1991qt,Klevansky:1992qe,Hatsuda:1994pi,Buballa} and the non-linear dynamics involving the Polyakov loop
\cite{Meisinger:1995ih,Meisinger:2001cq,Fukushima:2003fw,Mocsy:2003qw,Megias:2004hj}. In 
this Polyakov-loop-extended (PNJL) model, quarks develop quasiparticle masses by
propagating in the chiral condensate, while they couple at the same time to a
homogeneous background (temporal) gauge field representing Polyakov loop 
dynamics.

The ``classic'' NJL model incorporates the chiral symmetry of two-flavour QCD 
and its spontaneous breakdown at $T<T_c$. Gluonic degrees of freedom 
are ``integrated out'' and replaced by a local four-point interaction of
quark colour currents. Subsequent Fierz transformations project this 
interaction into various quark-antiquark and diquark channels. The colour
singlet $q{\bar q}$ modes of lowest mass are identified with the lightest
mesons. Pions properly emerge as Goldstone bosons at $T<T_c$. However, the
local $SU(N_c)$ gauge invariance of QCD is now replaced by a global $SU(N_c)$ symmetry
in the NJL model, so that the confinement property is lost. Consequently,
standard NJL-type models are bound to fail in attempts to describe $N_c=3$
thermodynamics around $T_c$ (and beyond) for non-zero quark chemical potential
$\mu$.

On the other hand the NJL model, with just the simplest possible one-parameter
colour current-current interaction between quarks, is remarkably successful
in reproducing the thermodynamics of $N_c=2$ Lattice QCD at finite 
$\mu$~\cite{Ratti:2004ra}. Encouraged by this result, the NJL quasiparticle
concept does suggest itself as a useful starting point. However, whereas aspects
of deconfinement are less significant in the $N_c=2$ case, they figure 
prominently for $N_c=3$. This motivates our extension towards the PNJL 
Lagrangian as a minimal approach incorporating both chiral symmetry restoration
and deconfinement.

The deconfinement phase transition is well defined in the heavy-quark limit,
where the Polyakov loop serves as an order parameter. This phase transition
is characterized by the spontaneous breaking of the $Z(3)$ center symmetry of
QCD \cite{Polyakov:1978vu,Susskind:1979up,Svetitsky:1982gs,Svetitsky:1985ye}. 
In the presence of dynamical quarks the center symmetry is explicitly broken.
No order parameter is established for the deconfinement
transition in this case~\cite{Fukushima:2002bk}, but the Polyakov loop
still serves as an indicator of a rapid crossover towards deconfinement. 
The chiral 
phase transition, on the other hand, has a well-defined order parameter in the
chiral limit of massless quarks: the chiral (or quark) condensate 
$\langle\bar{q}q\rangle$. This condensate, and its dynamical generation, is
the basic element of the original NJL model.

The primary aim of this paper is to test the effectiveness of the PNJL approach
when confronted with Lattice QCD thermodynamics. The PNJL Lagrangian is derived
in Section~\ref{sec2}. Parameters are fixed in Section~\ref{sec3} by
reproducing known properties of the pion and of the QCD vacuum in the hadronic phase, while the 
Polyakov loop effective potential is adjusted to pure gauge lattice results.
Sections~\ref{sec4} and~\ref{sec5} deal with the thermodynamics derived from 
the PNJL Lagrangian in the mean-field approximation. This discussion includes a
detailed comparison with Lattice QCD results at zero and at finite chemical potential.
\newpage

\section{The PNJL model\label{sec2}}
Following~\cite{Fukushima:2003fw} we introduce a generalized $N_f = 2$ Nambu 
Jona-Lasinio Lagrangian
with quarks coupled to a (spatially constant) temporal background gauge
field representing  Polyakov loop dynamics (the PNJL model):
\bea
\mathcal{L}_{PNJL}=\bar{\psi}\left(i\gamma_{\mu}D^{\mu}-\hat{m}_0
\right)\psi+\frac
{G}{2}\left[\left(\bar{\psi}\psi\right)^2+\left(\bar{\psi}i\gamma_5
\vec{\tau}\psi
\right)^2\right]
-\mathcal{U}\left(\Phi[A],\bar{\Phi}[A],T\right),
\label{lagr}
\eea
where $\psi=\left(\psi_u,\psi_d\right)^T$ is the quark field,
\bea
D^{\mu}=\partial^\mu-i A^\mu~~~~~~\mathrm{and}~~~~~~A^\mu=\delta_{\mu0}A^0~~.
\eea
The gauge coupling $g$ is conveniently absorbed in the definition of $A^\mu(x) = g {\cal A}^\mu_a(x){\lambda_a\over 2}$ where ${\cal A}^\mu_a$ is the SU(3) gauge field and $\lambda_a$ are the Gell-Mann matrices. The two-flavour current quark mass matrix is $\hat{m}_0 = diag(m_u, m_d)$ and we shall work in the isospin symmetric limit with $m_u = m_d \equiv m_0$. A local, chirally 
symmetric scalar-pseudoscalar four-point interaction of the quark fields is 
introduced with an effective coupling strength $G$. 

The quantity $\mathcal{U}(\Phi,\bar{\Phi},T)$ is the effective potential expressed in 
terms of the traced Polyakov loop\footnote{more precisely: the Polyakov line with periodic boundary conditions} and its (charge) conjugate,
\beq
\Phi=(\mathrm{Tr}_c\,L)/N_c,~~~~~~~~~~~~~~~~~~~~~\bar{\Phi}=(\mathrm{Tr}_c\,
L^{\dagger})/N_c.
\eeq
The Polyakov loop $L$ is a matrix in colour space explicitly given 
by
\beq
L\left(\vec{x}\right)=\mathcal{P}\exp\left[i\int_{0}^{\beta}d\tau\,A_4\left(\vec{x},\tau\right)\right],
\eeq
with $\beta = 1/T$ the inverse temperature and $A_4 = i A^0$. In a convenient gauge (the so-called
Polyakov gauge), the Polyakov loop matrix can be given a  
diagonal representation~\cite{Fukushima:2003fw}.

The coupling between Polyakov loop and quarks
is uniquely determined by the covariant derivative $D_\mu$ in the 
PNJL Lagrangian~(\ref{lagr}).
Note that in the chiral limit ($\hat{m}_0 \rightarrow 0$), this Lagrangian is 
invariant under the chiral flavour group, $SU(2)_L \times SU(2)_R$, just like 
the original QCD Lagrangian. 

The trace of the Polyakov loop, $\Phi$, and its conjugate, $\bar{\Phi}$, will 
be treated as classical field variables throughout this work.
In the absence of quarks, we have
$\Phi=\bar{\Phi}$ and the Polyakov
loop serves as an order parameter for deconfinement. The phase transition
is characterized by the spontaneous breaking of the $Z(3)$ center symmetry of 
QCD.
The temperature dependent effective potential $\mathcal{U}$ has the following
general features. 
At low 
temperatures, ${\cal U}$ has a single minimum at $\Phi$=0, while
at high
temperatures it develops a second one which turns into the absolute minimum
above a critical temperature $T_0$. In the limit $T\rightarrow\infty$ we have
$\Phi\rightarrow~1$. The function $\mathcal{U}(\Phi,\bar{\Phi},T)$ 
will be 
fixed by comparison with pure-gauge Lattice QCD. We choose the
following general form in accordance with the underlying $Z(3)$ symmetry:
\bea
{\mathcal{U}\left(\Phi,\bar{\Phi},T\right)\over T^4} =-{b_2\left(T\right)\over 2}\bar{\Phi} \Phi-
{b_3\over 6}\left(\Phi^3+
{\bar{\Phi}}^3\right)+ {b_4\over 4}\left(\bar{\Phi} \Phi\right)^2
\label{u1}
\eea
with
\beq
b_2\left(T\right)=a_0+a_1\left(\frac{T_0}{T}\right)+a_2\left(\frac{T_0}{T}
\right)^2+a_3\left(\frac{T_0}{T}\right)^3~~~.
\label{u2}
\eeq
A precision fit of the coefficients $a_i,~b_i$ is performed to reproduce the lattice data (see section 3).

Using standard bosonization techniques the 
Lagrangian~(\ref{lagr}) can be rewritten in terms of the auxiliary field variables $\sigma$ and
$\vec{\pi}$:
\bea
\mathcal{L}_{eff}=-\frac{\sigma^2+\vec{\pi}^2}{2G}
-\mathcal{U}\left(\Phi,\bar{\Phi},T\right)-i\mathrm{Tr}\ln
S^{-1},
\label{leff}
\eea
where an irrelevant constant has been dropped and
\beq
S^{-1}=i\gamma_{\mu}\partial^{\mu}-\gamma_0A^0-\hat{M}\, 
\eeq
is the inverse quark propagator with
\beq
\hat{M}=\hat{m}_0-\sigma-i\gamma_5\vec{\tau}\cdot\vec{\pi}\, .
\eeq
The trace in~(\ref{leff}) is taken over colour, flavour and Dirac indices. 
The field equations
for $\sigma$, $\vec {\pi}$, $\Phi$ and $\bar{\Phi}$
are then solved in the 
mean field approximation\footnote{In the mean field approximation the fields 
are replaced by their expectation values for which, in later sections, we will continue using the
notation $\sigma$, $\Phi$ and $\bar{\Phi}$ for simplicity and convenience.}. The 
expectation 
value 
$\langle\vec{\pi}\rangle$ of the pseudoscalar 
isotriplet field is equal to zero for isospin-symmetric systems. 

The 
$\sigma$ field has a 
non-vanishing vacuum expectation value as a 
consequence of spontaneous chiral symmetry breaking. Solving the field equations for $\sigma$, the 
effective quark mass $m$ is determined by the self-consistent gap equation
\beq
m=m_0-\langle\sigma\rangle=m_0-G\langle\bar{\psi}\psi\rangle.
\label{mass}
\eeq
Note that $\langle\sigma\rangle=G\langle\bar{\psi}\psi\rangle$ is negative in
our representation, and the chiral (quark) condensate is $\langle\bar{\psi}\psi\rangle=\langle\bar{\psi}_u
\psi_u+\bar{\psi}_d\psi_d\rangle$.
For later purposes we note that $\Phi$ and $\bar{\Phi}$ are two independent field 
variables in the general case of finite quark chemical potential $\mu$. They 
become equal in the limiting case $\mu=0$. Their (thermal) 
expectation values $\langle\Phi\rangle$ and $\langle\bar{\Phi}\rangle$ are both
real~\cite{Dumitru:2005ng} but differ at non-zero $\mu$.

Before passing to the actual calculations, we summarize basic 
assumptions behind eq.~(\ref{lagr}) and comment on limitations to be kept in 
mind. In fact
the PNJL model~(\ref{lagr}) is quite schematic in several respects. It reduces 
gluon dynamics to a) chiral point couplings between quarks, and b) a simple
static background field representing the Polyakov loop. This
picture cannot be expected to work beyond a limited range of temperatures.
At large $T$, transverse gluons are known to be thermodynamically active degrees
of freedom, but they are ignored in the PNJL model. To what extent this model
can reproduce lattice QCD thermodynamics is nonetheless a relevant question. We
can assume that its range of applicability is, roughly, $T\leq (2-3)T_c$, based
on the conclusion drawn in ref.~\cite{Meisinger:2003id} that transverse gluons
start to contribute significantly for $T>2.5\,T_c$.

\section{Parameter fixing\label{sec3}}
\subsection{Polyakov loop effective potential}
\begin{figure}
\parbox{5cm}{
\scalebox{1}{
\includegraphics*[19,502][381,730]{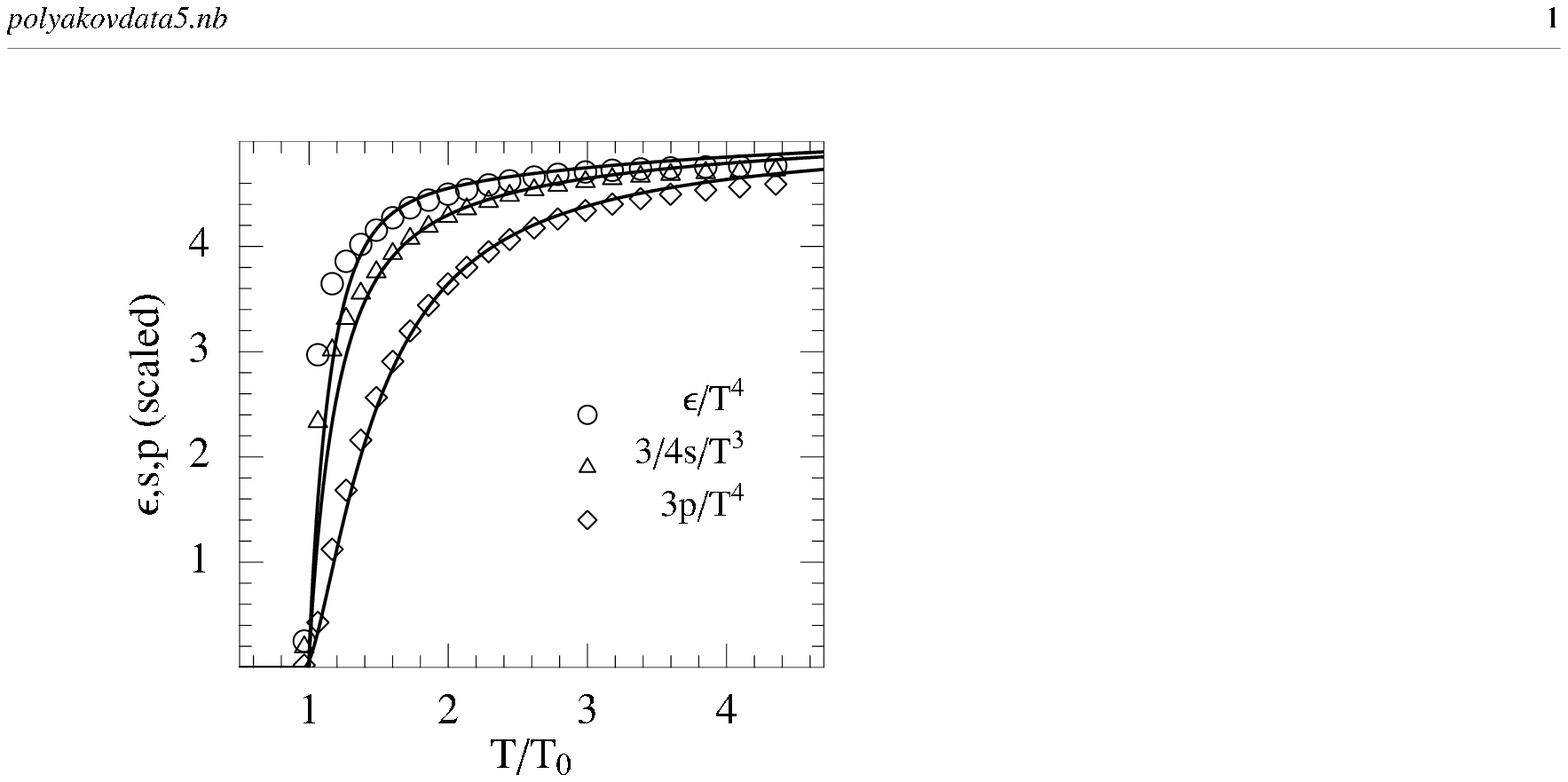}}}\\
\parbox{15cm}{
\caption{
\footnotesize Scaled pressure, entropy density and energy density as functions
of the temperature in the pure gauge sector, compared to the corresponding 
lattice data taken from Ref.~\cite{Boyd}.
\label{fig1}}}
\end{figure}

\begin{figure}
\hspace{-3cm}
\begin{minipage}[t]{.48\textwidth}
\parbox{5cm}{
\scalebox{1.05}{
\includegraphics*[width=\textwidth]{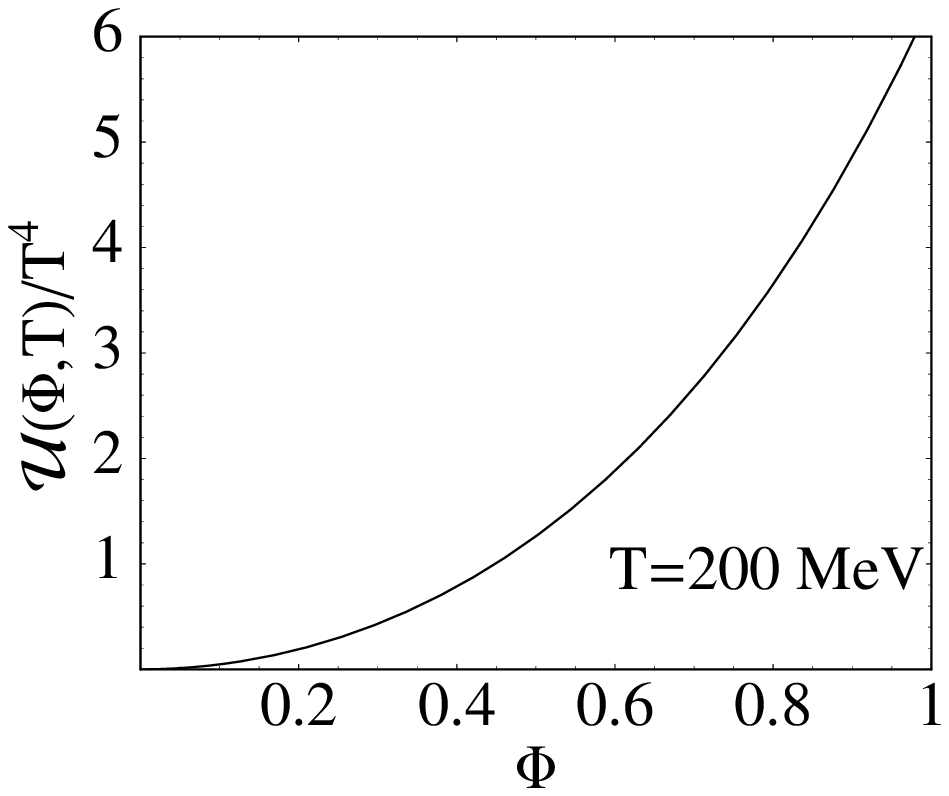}\\}}
\end{minipage}
\hspace{-3cm}
\begin{minipage}[t]{.48\textwidth}
\parbox{5cm}{
\scalebox{1.05}{
\includegraphics*[width=\textwidth]{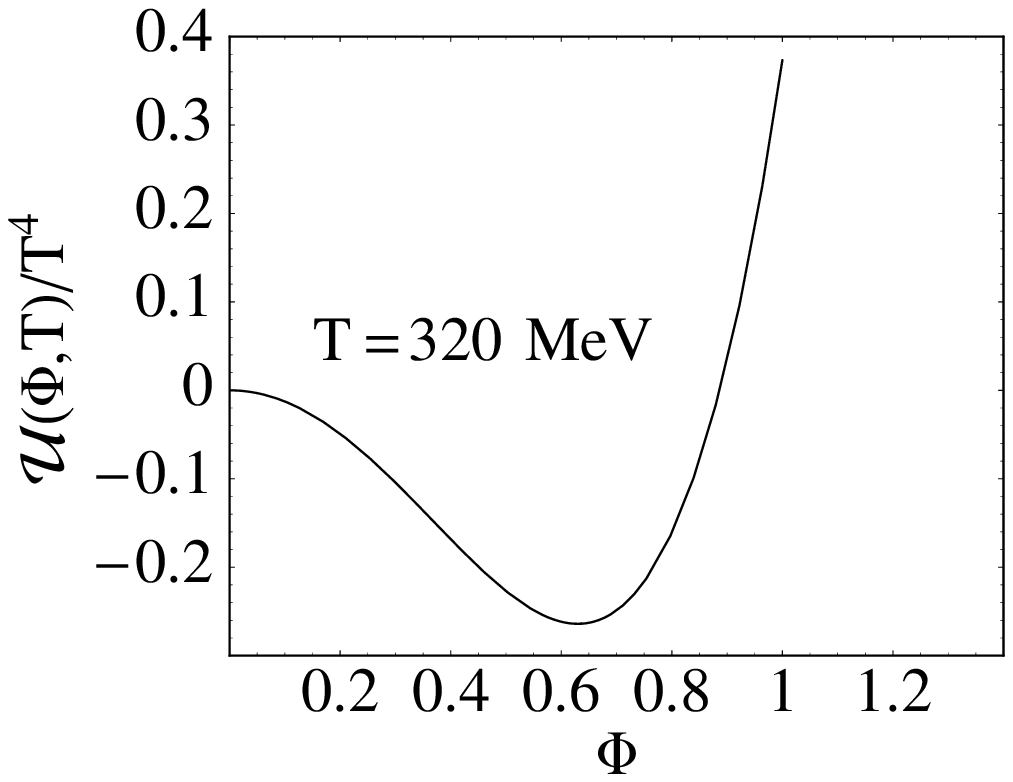}\\}}
\end{minipage}
\parbox{15cm}{
\caption{
\footnotesize 
Scaled Polyakov loop effective potential $\mathcal{U}(\Phi,T)/T^4$ as a function of
$\Phi$ for two values of temperature $T$.}
\label{fig:efp}
}
\end{figure}

The parameters of the Polyakov loop potential $\mathcal{U}$ are fitted to reproduce 
the lattice data~\cite{Boyd} for QCD thermodynamics in the pure gauge sector.
Minimizing $\mathcal{U}(\Phi,\bar{\Phi},T)$ one has $\Phi=\bar{\Phi}$ and the pressure
of the pure-gauge system is evaluated as $p(T)=-\mathcal{U}(T)$ with $\Phi(T)$ 
determined at the minimum.
The entropy and energy density are then obtained by means of the standard
thermodynamic relations.
In Fig.~\ref{fig1} we 
show the (scaled) pressure, energy density and entropy density as functions of 
temperature. The lattice data are
reproduced extremely well using the ansatz~(\ref{u1}, \ref{u2}), with 
parameters summarized in Table~\ref{table1}.
The critical temperature $T_0$ for deconfinement appearing in Eq.~(\ref{u2}) is 
fixed at $T_0=270$ MeV in the pure gauge sector. The resulting effective potential 
is displayed in Fig.~\ref{fig:efp} for two different temperatures:
$T =200$ MeV (below $T_0$) and $T=320$ MeV (above $T_0$). 

\begin{table}
\begin{center}
\begin{tabular}{|c|c|c|c|c|c|}
\hline
\hline
&&&&&\vspace{-.3 cm}\\
$a_0$&$a_1$&$a_2$&$a_3$&$b_3$&$b_4$\\
&&&&&
\vspace{-.3cm}\\
\hline
\vspace{-.3cm}\\
6.75&-1.95&2.625&-7.44&0.75&7.5\\
\hline
\hline
\end{tabular}
\caption{Parameter set used in this work for the Polyakov loop 
potential~(\ref{u1},~\ref{u2}).}
\label{table1}
\end{center}
\end{table}

\begin{figure}
\parbox{5cm}{
\scalebox{1.1}{
\includegraphics*[39,522][381,730]{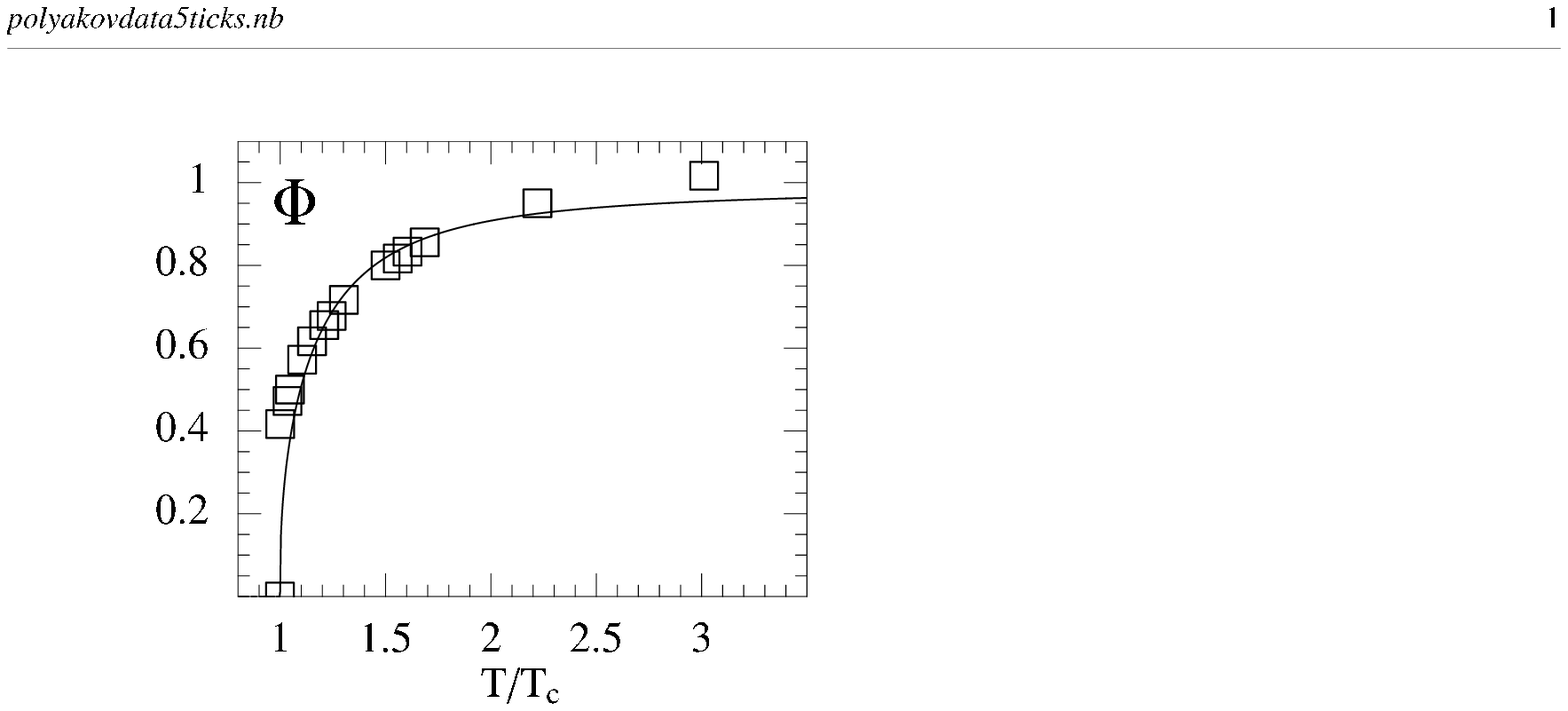}}
}\\
\parbox{15cm}{
\caption{
\footnotesize Polyakov loop as a function of  
temperature in the pure gauge sector, compared to corresponding lattice results taken from Ref.~\cite{Kaczmarek:2002mc}.
\label{fig1b}}
}
\end{figure}

With the same parametrization, we are also able to reproduce the 
lattice data~\cite{Kaczmarek:2002mc} for the temperature dependence of the Polyakov 
loop itself.  A comparison between these data and our results is shown
in Fig.~\ref{fig1b}. The Polyakov loop vanishes below the critical temperature $T_0$, at which point
it jumps discontinuously to a finite value, indicating a first order phase 
transition. It tends to one at large 
temperatures, as expected.

\subsection{NJL sector}

The pure NJL model part of the Lagrangian~(\ref{lagr}) has the following parameters: the ``bare'' quark mass 
$m_0$, 
a three-momentum cutoff $\Lambda$ and the coupling strength $G$.
We choose to reproduce the known chiral physics in the hadronic sector at $T=0$
and fix 
the three parameters by the following conditions:
\begin{itemize}
\item{The pion decay constant is reproduced at its empirical value, $f_{\pi}=92.4$ MeV. 
In the NJL model, $f_\pi$ is evaluated using the following equation:
\beq
f_{\pi}^{2}=4m^2I_{\Lambda}^{(1)}\left(m\right)~~~\mathrm{where}~~~
I_{\Lambda}^{(1)}\left(m\right)
=-iN_c\int\frac{d^4p}{\left(2\pi\right)^4}\frac{\theta\left(\Lambda^2-
\vec{p}^{~2}
\right)}{\left(p^2-m^2+i\epsilon\right)^2} \, ,
\label{fp}
\eeq
with the effective (constituent) quark mass $m$ determined self-consistently by the gap equation (\ref{mass}).}
\item{The quark condensate becomes
\beq
\left<{\bar \psi}_u\psi_u\right>=
-4m~I_{\Lambda}^{(0)}\left(m\right)
\eeq
with
\beq
I_{\Lambda}^{(0)}\left(m\right)=iN_c\int\frac{d^4p}{\left(2\pi\right)^4}\frac
{\theta\left(\Lambda^2-\vec{p}^{~2}\right)}{p^2-m^2+i\epsilon}\, .
\eeq
Its ``empirical'' value derived from QCD sum rules is 
\beq
\langle\bar{\psi}_u\psi_u\rangle^{1/3}\simeq\langle\bar{\psi}_d\psi_d\rangle
^{1/3}=-\left(240\pm20\right)~\mathrm{MeV}\, .
\eeq
}
\item{The current quark mass $m_0$ is fixed from the Gell-Mann, Oakes, Renner
(GMOR) relation which is satisfied in the NJL model:
\beq
m_{\pi}^{2}=\frac{-m_0\left<{\bar\psi}\psi\right>}{f_{\pi}^2}\,\,\, .
\label{gmor}
\eeq
In the chiral limit, $m_0=0$ and $m_{\pi}=0$.}
\end{itemize}


The values of the NJL model parameters, together with the resulting physical
quantities, are summarized in Table~\ref{table2}.

\begin{table}
\begin{center}
\begin{tabular}{|c|c|c|}
\hline
\hline
&&\vspace{-.3 cm}\\
$\Lambda$ [GeV]&$G$[GeV$^{-2}$]&$m_0$[MeV]\\
&&\vspace{-.3cm}\\
\hline
&&\vspace{-.3cm}\\
0.651&10.08&5.5\\
&&\vspace{-.3cm}\\
\hline
\hline
&&\vspace{-.3cm}\\
$|\langle{\bar \psi}_u\psi_u\rangle|$$^{1/3}$[MeV]&$f_{\pi}$[MeV]&$m_{\pi}$[MeV]\\
&&\vspace{-.3cm}\\
\hline
&&\vspace{-.3cm}\\
251&92.3&139.3\\
\hline
\hline
\end{tabular}
\caption{Parameter set used in this work for the NJL model part of the effective Lagrangian~(\ref{lagr}),
and the resulting physical quantities. For these values of the parameters we
obtain a constituent quark mass $m=$325 MeV.}
\label{table2}
\end{center}
\end{table}

\section{Results at finite $T$ and $\mu$\label{sec4}}

We now extend the model to finite temperature and
chemical potentials using the Matsubara formalism. We
consider the isospin symmetric case, with an equal number of $u$ and $d$ quarks (and therefore a 
single quark chemical potential $\mu$).
The quantity to be minimized at finite temperature is the thermodynamic
potential per unit volume:
\bea
\Omega\left(T,\mu\right)&=&{\cal U}\left(\Phi,\bar{\Phi},T\right)-{T\over 2}\sum_n\int\frac{d^3p}{\left(2\pi\right)^3}
\mathrm{Tr}\ln{\tilde{S}^{-1}\left(i\omega_n,\vec{p}\,\right)\over T}
+\frac{\sigma^2}{2G}.\nonumber\\
\label{omega}
\eea
Here $\omega_n=(2n+1)\pi T$ are the Matsubara frequencies for fermions. The inverse quark propagator (in Nambu-Gorkov representation) becomes 
\beq
\tilde{S}^{-1}\left(p^0,\vec{p}\,\right)=\left({{\begin{array}{ccc} 
\gamma_0p^0-\vec{\gamma}\cdot\vec{p}-
m-\gamma_0\left(\mu + i A_4\right)& 0\\0& 
\gamma_0p^0-\vec{\gamma}\cdot\vec{p}-
m+\gamma_0\left(\mu + i A_4\right)
\end{array}}}\right).
\label{propmu}
\eeq
Using the identity $\mathrm{Tr}\ln\left(X\right)=\ln\det\left(X\right)$
we reduce the trace in (\ref{omega}) and find:
\bea
\nonumber
\Omega&=&{\cal U}\left(\Phi,\bar{\Phi},T\right)+\frac{\sigma^2}{2G}\\
&-&2N_f\,T\int\frac{\mathrm{d}^3p}{\left(2\pi\right)^3}
\left\{\mathrm{Tr}_c\ln\left[1+L\,\mathrm{e}^{-\left(E_p-\mu
\right)/T}\right]
+\mathrm{Tr}_c\ln\left[1+L^{\dagger}\,\mathrm{e}^{-\left(E_p+
\mu\right)/T}\right]\right\}
\nonumber\\ 
&-& 6N_f\int\frac{\mathrm{d}^3p}{\left(2\pi\right)^3}{E_p}\,\theta\left(\Lambda^2-\vec{p}^{~2}\right)\nonumber\\
\label{omega2}
\eea
where we have introduced the quark quasiparticle energy 
$E_p=\sqrt{\vec{p}^{~2}+m^2}$. The last term involves the NJL three-momentum 
cutoff $\Lambda$. The second (finite) term does not require any cutoff. A small
violation of the underlying chiral symmetry at $T>0.4$ GeV, resulting from this
procedure, is of no practical relevance since the model is supposed to be applied only at temperatures
and chemical potential well below $\Lambda$.

The remaining colour trace is then performed with the result 
\bea
\nonumber
&&\ln \mathrm{det}\left[1+L\,\mathrm{e}^{-\left(E_p-\mu\right)/T}\right]+
\ln \mathrm{det}\left[1+L^{\dagger}\,\mathrm{e}^{-\left(E_p+
\mu\right)/T}\right]\\
&=&\ln\left[1+3\left(\Phi+\bar{\Phi}\mathrm{e}^{-\left(E_p-\mu\right)/T}\right)\mathrm{e}^{-\left(E_p-\mu\right)/T}+
\mathrm{e}^{-3\left(E_p-\mu\right)/T}\right]
\nonumber\\
&+&
\ln\left[1+3\left(\bar{\Phi}+\Phi\mathrm{e}^{-\left(E_p+\mu\right)/T}\right)\mathrm{e}^{-\left(E_p+\mu\right)/T}+
\mathrm{e}^{-3\left(E_p+\mu\right)/T}\right].
\eea
From the thermodynamic potential (\ref{omega2}) the equations of motion for the mean fields $\sigma, \Phi$ and $\bar{\Phi}$ are derived through
\beq
{\partial\Omega\over\partial\sigma} = 0 ~,~~~~~{\partial\Omega\over\partial\Phi} = 0 ~,~~~~~{\partial\Omega\over\partial \bar{\Phi}} = 0 ~.
\eeq
This set of coupled equations is then solved for the fields as functions of temperature $T$ and quark chemical potential $\mu$.
\begin{figure}
\hspace{-.05\textwidth}
\begin{minipage}[t]{.48\textwidth}
\includegraphics*[width=\textwidth]{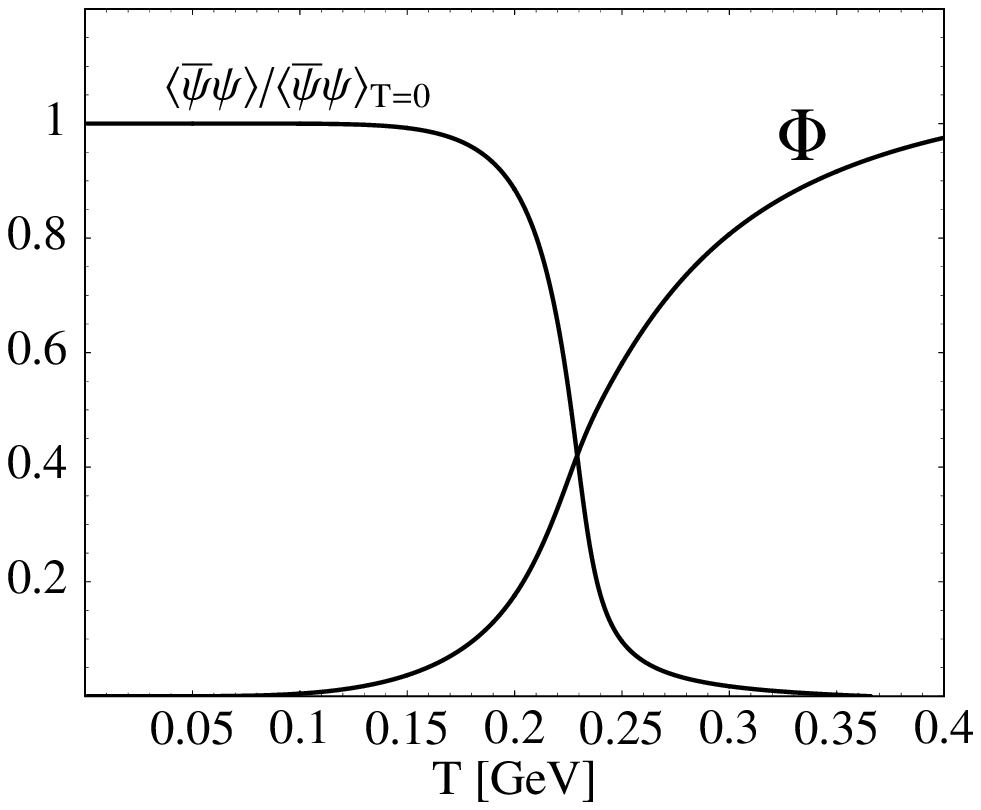}\\
\centerline{(a)}
\end{minipage}
\hspace{-.05\textwidth}
%
\begin{minipage}[t]{.48\textwidth}
\hspace{-.15\textwidth}
\scalebox{.85}{
\includegraphics*[width=1.17\textwidth]{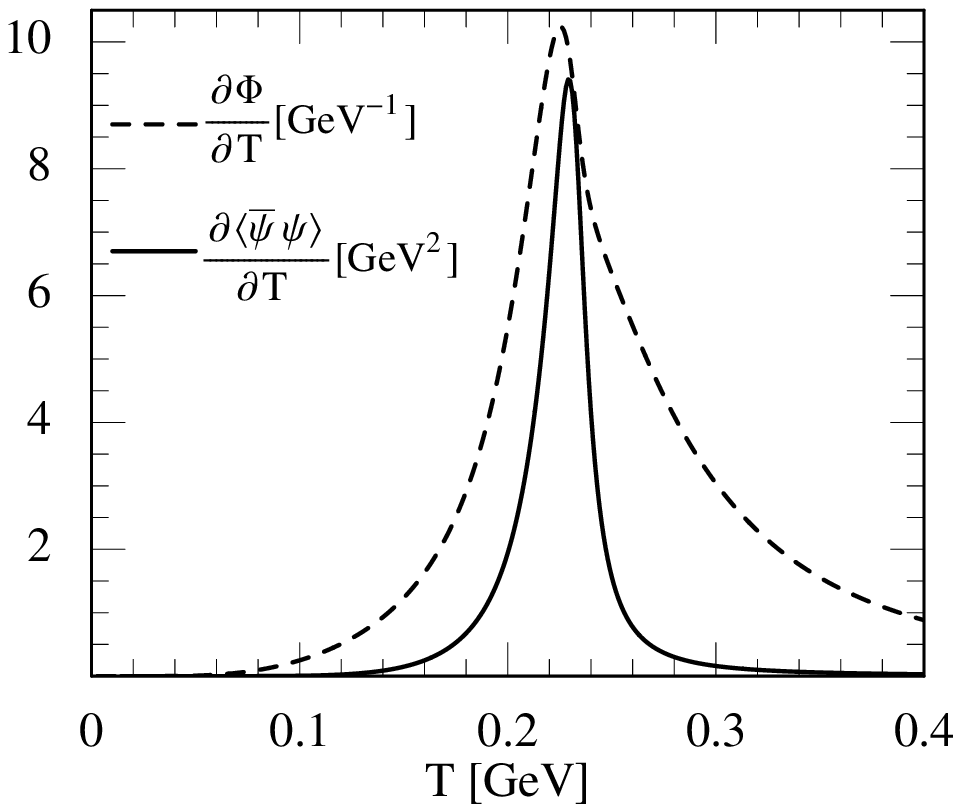}}\\
\centerline{(b)}
\end{minipage}
\parbox{15cm}{
\caption{Left: scaled chiral condensate and Polyakov loop 
$\Phi(T)$ as
functions of temperature at zero chemical potential. Right: plots of
$\partial\langle\bar{\psi}\psi\rangle/\partial T$ and $\partial\Phi/
\partial T$.
}
\label{fig2}}
\end{figure}
%
%

Fig.~\ref{fig2}(a) shows the chiral condensate together with the Polyakov loop 
$\Phi$ as functions of temperature at 
$\mu = 0$ where we find again $\Phi=\bar{\Phi}$.
One observes that the introduction of quarks coupled to the $\sigma$ and
$\Phi$ fields turns the first-order transition seen in pure-gauge Lattice QCD
into a continuous crossover. The original 1st order transition in the
pure-gauge system appears at a critical temperature $T_0 = 270$ MeV. With
the introduction of quarks, the 
crossover transitions for the chiral condensate 
$\langle\bar{\psi}\psi\rangle$ and for the Polyakov loop perfectly coincide 
at a lower critical temperature $T_c \simeq 220$ MeV (see Fig.~\ref{fig2}(b)). 
We point out that this feature is obtained without changing a single
parameter with respect to the pure gauge case.
The value of the critical temperature that we obtain is a little high if 
compared to the available data for two-flavour Lattice QCD 
\cite{Karsch} which gives $T_c = (173 \pm 8)$ MeV. On the other hand, it is
presently being discussed that detailed continuum extrapolation of these data  
can increase this temperature 
up to 210 MeV~\cite{Fodor}.
For quantitative comparison with existing lattice results we choose to reduce 
$T_c$ by rescaling the parameter $T_0$ from 270 to 190 MeV. In this case we 
loose the perfect
coincidence of the chiral and deconfinement transitions, but they are shifted
relative to each other by less than 20 MeV. When defining $T_c$ in this case
as the average of the two transition temperatures we find $T_c=180$ MeV.
This is also consistent with the observations reported in \cite{Digal}.

\begin{figure}
\begin{center}
\includegraphics*[width=.6\textwidth]{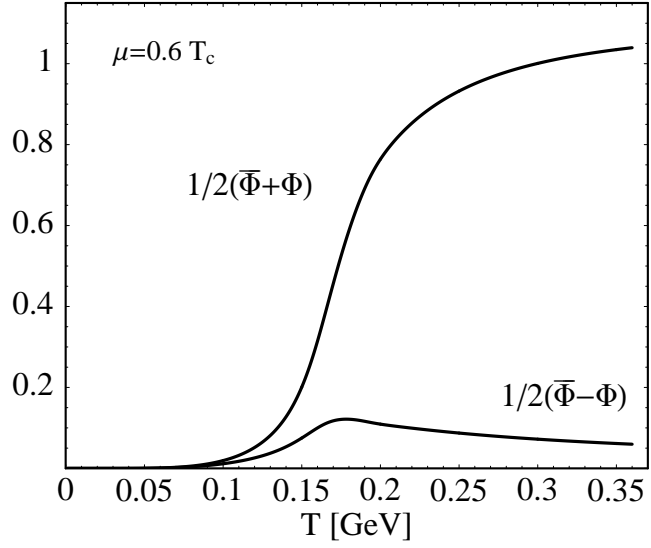}\\
\end{center}
\parbox{15cm}{
\caption{
\footnotesize Averaged sum and difference of $\Phi$ and $\bar{\Phi}$ as functions of the temperature at finite $\mu$.
\label{sumdifference}}
}
\end{figure}

As we turn to non-zero chemical potential, we find that $\Phi$ and
$\bar{\Phi}$ are different from each other, even if they are both
real. They will finally 
coincide again at high temperatures, as can be seen in 
Fig.~\ref{sumdifference}.
This feature was already observed in ~\cite{Dumitru:2005ng}.

\begin{figure}
\hspace{-.05\textwidth}
\begin{minipage}[t]{.48\textwidth}
\parbox{5cm}{
\scalebox{.95}{
\includegraphics*[width=\textwidth]{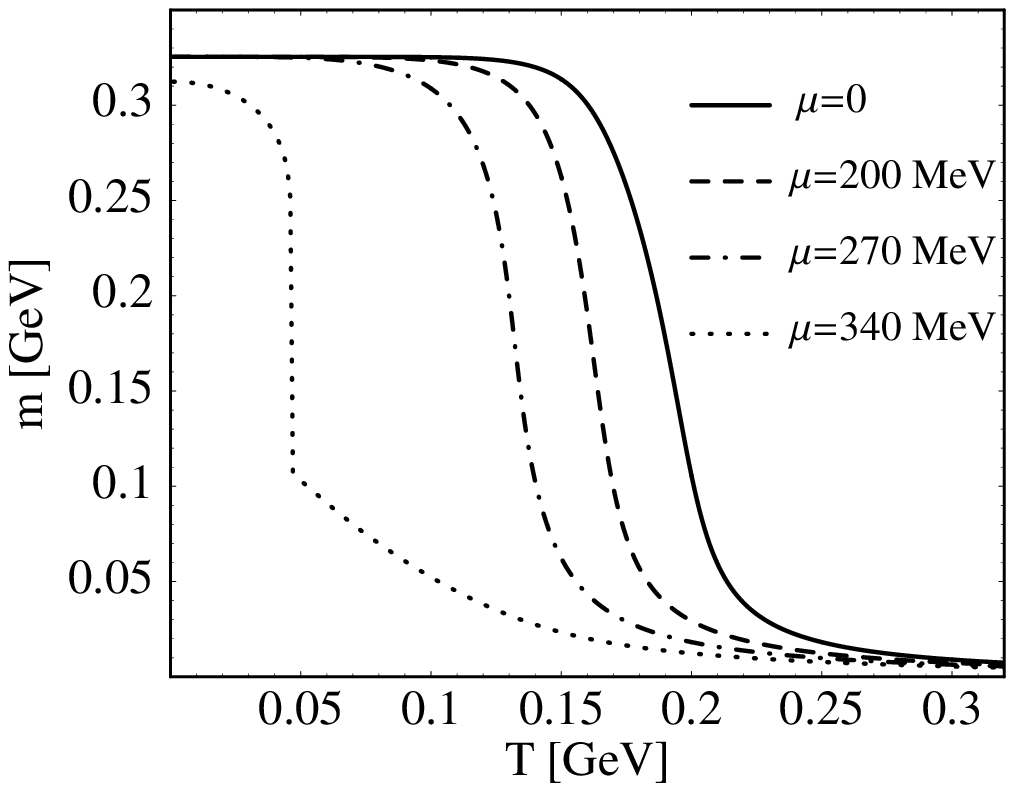}\\}}
\centerline{(a)}
\end{minipage}
\hspace{.02\textwidth}
\begin{minipage}[t]{.48\textwidth}
\parbox{5cm}{
\scalebox{0.95}{
\includegraphics*[width=\textwidth]{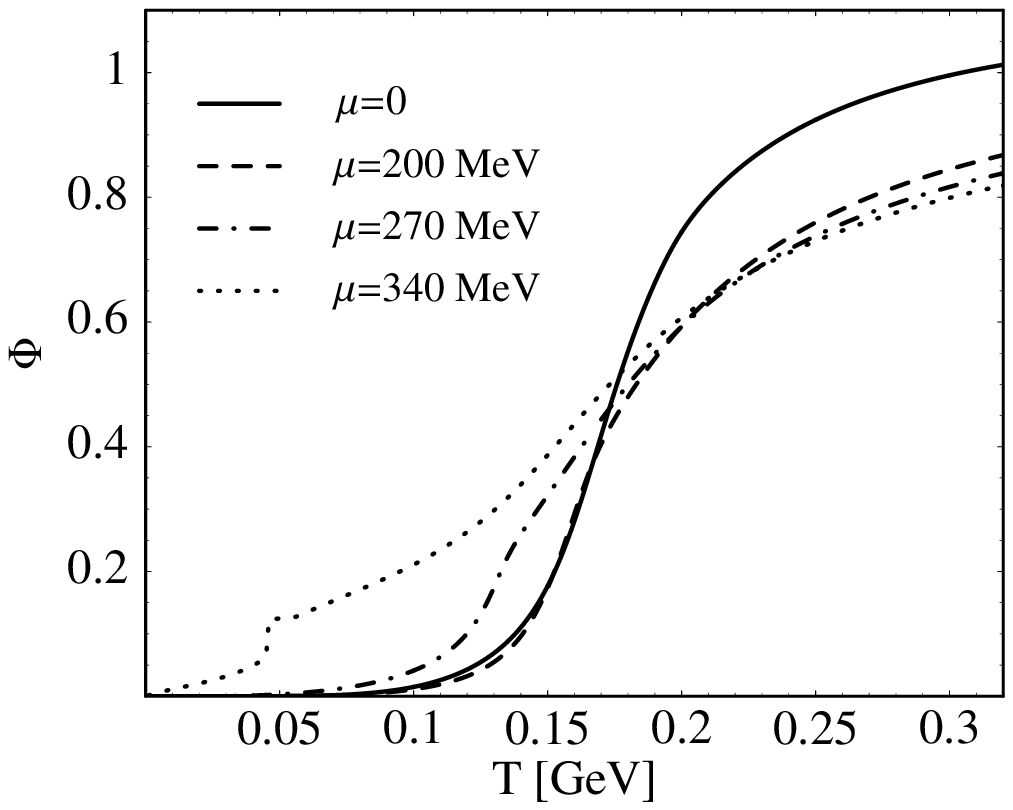}\\}}
\centerline{(b)}
\end{minipage}
\parbox{15cm}{
\caption{
\footnotesize Constituent quark mass $(a)$ and Polyakov loop
$(b)$ as functions of temperature for different values of the chemical 
potential.}
\label{figqmpl}}
\end{figure}

With increasing chemical potential, the crossover pattern evolves to lower 
transition temperatures (see Fig.~\ref{figqmpl}) until it turns to a first 
order transition around $\mu \sim 0.3$ GeV. At this point Cooper pairing of 
quarks presumably sets in. A more detailed discussion of the critical point and
its neighbourhood therefore requires the additional incorporation of explicit 
diquark degrees of freedom in the PNJL model. Further developments along these 
lines will be reported elsewhere. 
\section{Detailed comparison with Lattice QCD\label{sec5}}

\begin{figure}
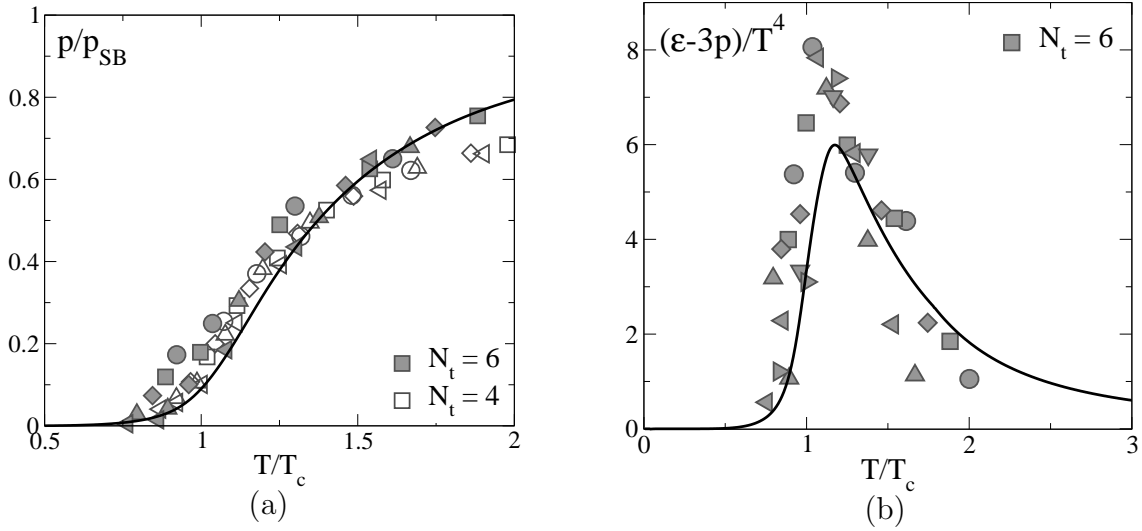

\hspace{-.05\textwidth}
\begin{minipage}[t]{.48\textwidth}
\parbox{5cm}{
\scalebox{.95}{
\includegraphics*[width=\textwidth]{fig7a.eps}\\}}
\centerline{(a)}
\end{minipage}
\hspace{.02\textwidth}
\begin{minipage}[t]{.48\textwidth}
\parbox{5cm}{
\scalebox{0.95}{
\includegraphics*[width=\textwidth]{fig7b.eps}\\}}
\centerline{(b)}
\end{minipage}
\parbox{15cm}{
\caption{
\footnotesize (a) Scaled pressure divided by the Stefan-Boltzmann (ideal gas) 
limit as a function of temperature at zero chemical
potential: comparison between our PNJL model prediction and 
lattice results corresponding to $N_t=4$ and $N_t=6$. (b) Scaled interaction 
measure compared to lattice results for $N_t=6$.
In both cases, the lattice data are taken from Ref.~\cite{AliKhan:2001ek}}
\label{fig4}}
\end{figure}

A primary aim of this work is to compare predictions of our PNJL model with the
lattice data available for full QCD thermodynamics at zero and finite 
$\mu$. Consider first the pressure of the quark-gluon
system at zero chemical potential:
\beq
p\left(T,\mu=0\right)=-\Omega\left(T,\mu=0;\sigma(T,0),\Phi(T,0),\bar{\Phi}(T,0)\right),
\eeq
where $\sigma\left(T,0\right),~\Phi\left(T,0\right)$ and $\bar{\Phi}\left(T,0\right)$ are the solutions of the 
field equations at finite temperature and zero quark chemical potential.
Our results are presented in Fig.~\ref{fig4}(a) in comparison with 
corresponding lattice data. We point out that the input parameters
of the PNJL model have been fixed independently in the pure gauge and hadronic sectors, so that
our calculated pressure is a prediction of the model, without any further tuning of parameters. 
With this in mind, the agreement with lattice results is quite satisfactory. One must note that the lattice data are grouped in different sets obtained on lattices with temporal extent $N_t = 4$ and $N_t = 6$, both of which are not continuum extrapolated. In contrast, our calculation should, strictly speaking, be compared to the continuum limit. In order to perform meaningful comparisons, the pressure is divided by its asymptotic high-temperature (Stefan-Boltzmann) limit for each given case. At high temperatures our predicted curve should be located closer to the $N_t = 6$ set than to the one with $N_t = 4$. This is indeed the case. Furthermore, Fig.~\ref{fig4}(b) shows the predicted ``interaction measure", $(\varepsilon - 3p)/T^4$, in comparison with lattice data for $N_t = 6$. One should of course note that the lattice results have been produced using relatively large quark masses, with pseudoscalar-to-vector mass ratios $m_{PS}/m_V$ around 0.7, whereas our calculation is performed with light quark masses corresponding to the physical pion mass. We have investigated the dependence of the pressure and of the energy density on the quark mass and found that the critical temperature scales approximately as $T_c \simeq T_c(m_\pi = 0) + 0.04\,m_\pi$, in agreement with the behaviour found in \cite{Karsch}. Once the rescaling of $T_c$ is taken into account, the curves plotted in Figs.~\ref{fig4}(a) and \ref{fig4}(b) as functions of $T/T_c$ have negligible remaining dependence on the quark mass. 

\begin{figure}
\hspace{-.05\textwidth}
\begin{minipage}[t]{.48\textwidth}
\includegraphics*[width=\textwidth]{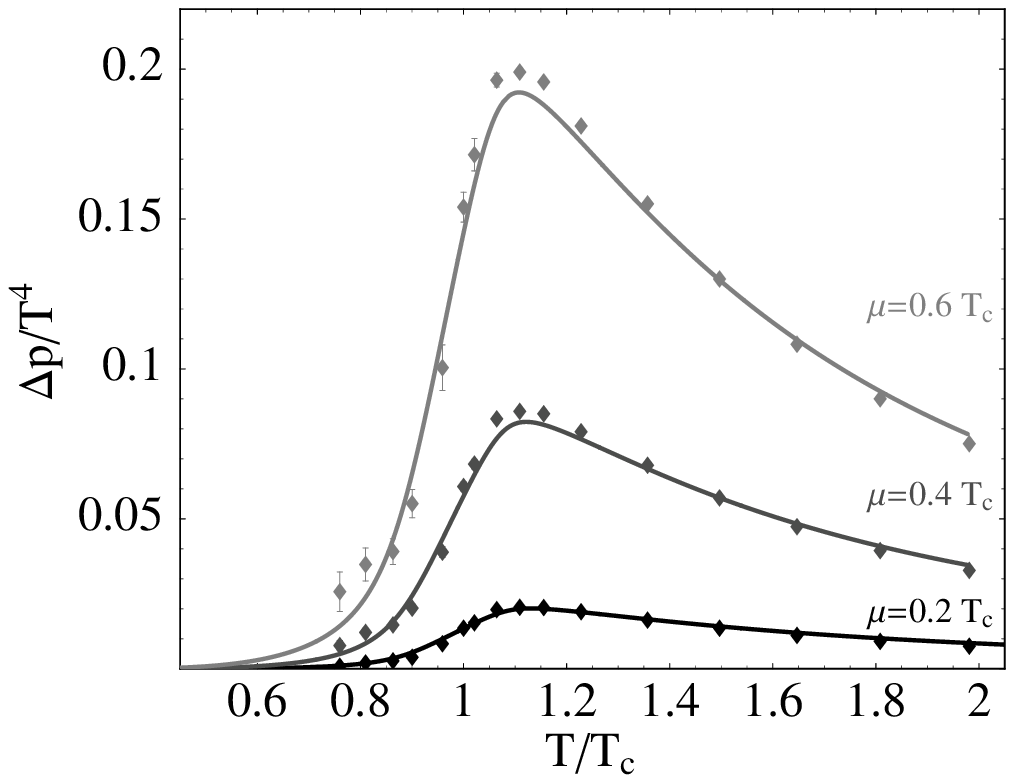}\\
\centerline{(a)}
\end{minipage}
\hspace{.02\textwidth}
\begin{minipage}[t]{.48\textwidth}
\includegraphics*[width=\textwidth]{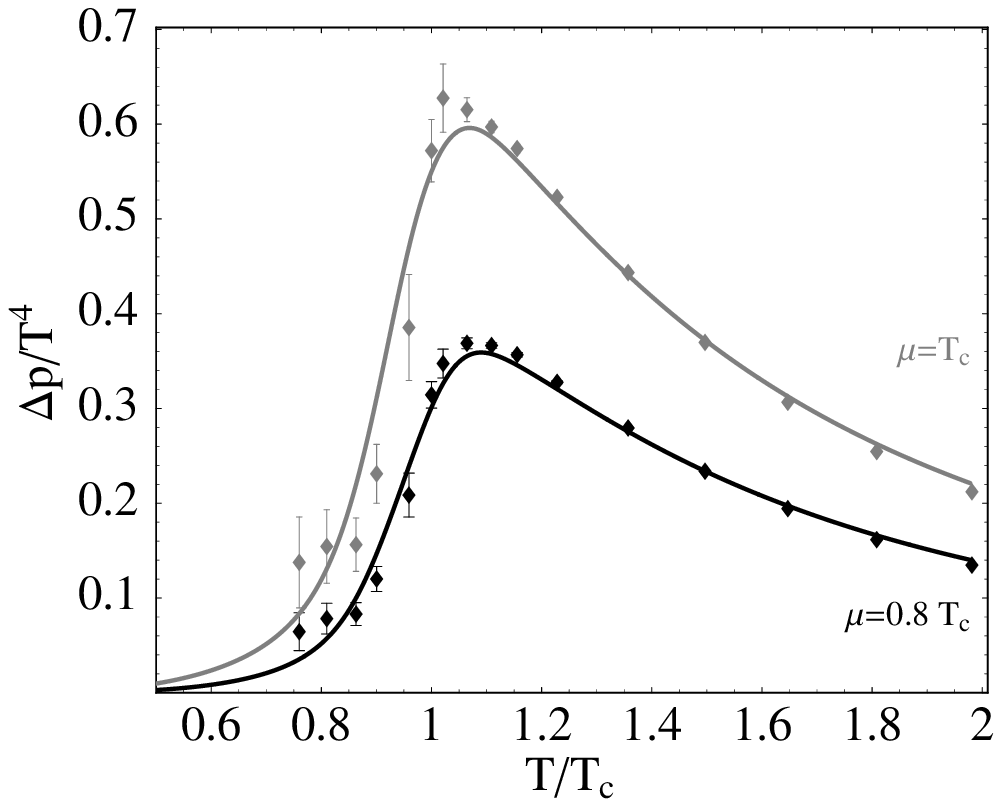}\\
\centerline{(b)}
\end{minipage}
\parbox{15cm}{
\caption{
\footnotesize Scaled pressure difference as a function of temperature at
different values of the quark chemical potential, compared to 
lattice data taken from Ref.~\cite{Allton:2003vx}.}
\label{fig5}}
\end{figure}

At non-zero chemical potential, quantities of interest that have become accessible in Lattice QCD are the ``pressure difference" and the quark number density.
The (scaled) pressure difference is defined as:
\bea
\frac{\Delta p\left(T,\mu\right)}{T^4}=\frac{p\left(T,\mu\right)-p\left(T,
\mu=0\right)}{T^4}.
\eea
A comparison of $\Delta p$, calculated in the PNJL model, with lattice results is presented in Fig.~\ref{fig5}. This figure shows
the scaled pressure difference as a function of the temperature for a series of
chemical potentials, with values ranging between $\mu=0.2\,T_{c}^{(0)}$ and
$\mu \simeq T_{c}^{(0)}$. The agreement between our results and the lattice 
data is quite satisfactory.

\begin{figure}
\hspace{-.05\textwidth}
\begin{minipage}[t]{.48\textwidth}
\includegraphics*[width=\textwidth]{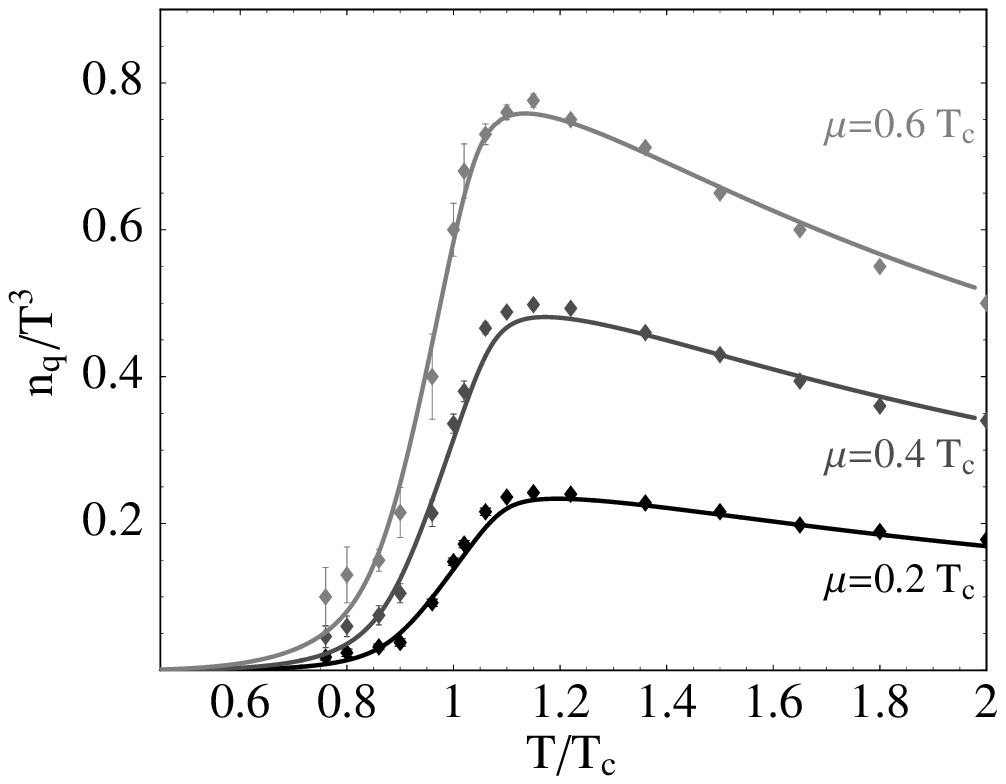}\\
\centerline{(a)}
\end{minipage}
\hspace{.02\textwidth}
\begin{minipage}[t]{.48\textwidth}
\includegraphics*[width=\textwidth]{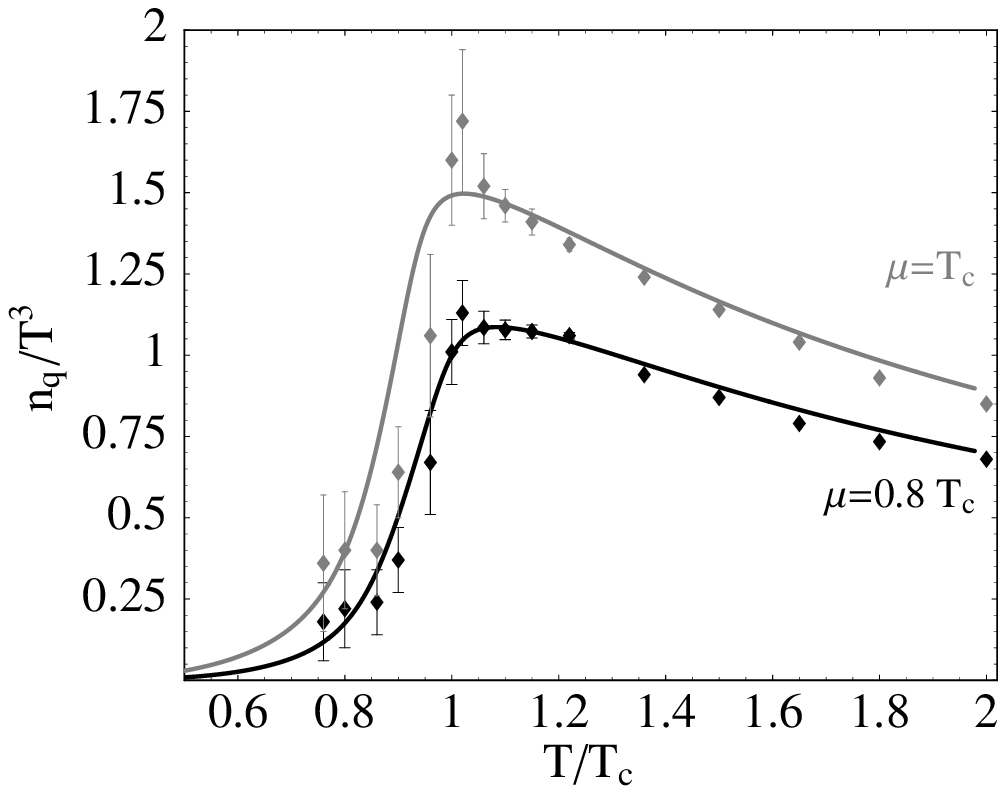}\\
\centerline{(b)}
\end{minipage}
\parbox{15cm}{
\caption{
\footnotesize Scaled quark number densities as a function of temperature at
different values of the chemical potential, compared to 
lattice data taken from Ref.~\cite{Allton:2003vx}.}
\label{fig6}}
\end{figure}

A related quantity for which lattice results at finite $\mu$ exist, is
the scaled quark number density, defined as:
\beq
\frac{n_q\left(T,\mu\right)}{T^3}=-\frac{1}{T^3}\frac{\partial\Omega
\left(T,\mu\right)}{\partial\mu}.
\eeq
Our results for $n_q$ as a function of the temperature,
for different values of the quark chemical potential, are shown in Fig.~\ref{fig6}
in comparison with corresponding lattice data \cite{Allton:2003vx}. Also in this case, the 
agreement between our PNJL model  and the corresponding lattice
data is surprisingly good. 
\begin{center}
\hspace{3cm}
\begin{figure}
\parbox{5cm}{
\scalebox{1}{
\hspace{4cm}
\centerline{\includegraphics*[width=.5\textwidth]{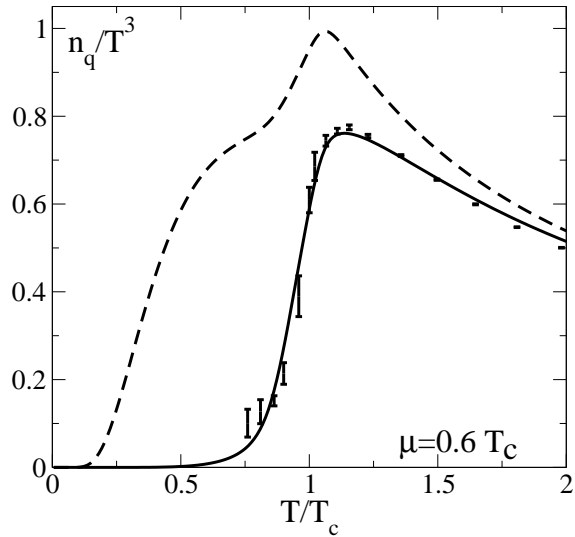}}}}\\
\parbox{15cm}{
\caption{
\footnotesize Comparison between the results in the PNJL model (solid
line) and in the standard NJL model (dashed line) for the quark number density
at $\mu=0.6 T_c$. The effect of the missing confinement is evident in the 
standard NJL model. 
\label{fig3}}
}
\end{figure}
\end{center}
It is instructive to study the effect of the Polyakov loop dynamics on the 
behaviour of the quark density $n_q$. The coupling of the quark quasiparticles 
to the field $\Phi$ reduces their weight as thermodynamically active degrees of
freedom when the critical temperature $T_c$ is approached from above. At $T_c$ 
the value of $\Phi$ tends to zero and the quasiparticle exponentials
exp$[-(E_p \pm \mu)/T]$ are progressively suppressed in the thermodynamic 
potential as $T \rightarrow T_c$. This is what can be interpreted as the impact
of confinement in the context of the PNJL model. In contrast, the standard NJL
model without coupling to the Polyakov loop does not have this important 
feature, so that the quark density leaks strongly into the ``forbidden" domain 
$T < T_c \simeq 170$ MeV, as demonstrated in Fig.~\ref{fig3}. 

It is a remarkable feature that the quark densities and the pressure difference at finite $\mu$ are
so well reproduced even though the lattice ``data'' have been obtained by a Taylor expansion up to fourth
order in $\mu$, whereas our thermodynamic potential is used with its full functional dependence on $\mu$. We have examined the convergence in powers of $\mu$ by expanding Eq.(\ref{omega2}). It turns out that the Taylor expansion to order $\mu^2$ deviates from the full result by less than 10 \% even at a chemical potential as large as $\mu\sim T_c$. When expanded to ${\cal O}(\mu^4)$,
no visible difference is left between the approximate and full calculations for all cases shown in Figs. 8 and 9.

\section{Summary and conclusions}
We have studied a Polyakov-loop-extended Nambu and Jona-Lasinio (PNJL) model with the aim of exploring whether such an approach can catch essential features of QCD thermodynamics when confronted with results of lattice computations at finite temperature and non-zero quark chemical potential. This PNJL model represents a minimal synthesis of the two basic principles that govern QCD at low temperatures: spontaneous chiral symmetry breaking and confinement. The  respective order parameters (the chiral quark condensate and the Polyakov loop) are given the meaning of collective degrees of freedom. Quarks couple to these collective fields according to the symmetry rules dictated by QCD itself.

Once a limited set of input parameters is fitted to Lattice QCD in the pure gauge sector and to pion properties in the hadron sector, the quark-gluon thermodynamics above $T_c$ up to about twice the critical temperature is well reproduced, including quark densities up to chemical potentials of about 0.2 GeV. In particular, the PNJL model correctly describes the step from the first-order deconfinement transition observed in pure-gauge Lattice QCD (with $T_c \simeq 270$ MeV) to the crossover transition (with $T_c$ around 200 MeV) when $N_f = 2$ light quark flavours are added. The non-trivial result is that the crossovers for chiral symmetry restoration and deconfinement almost coincide, as found in lattice simulations. The model also reproduces the quark number densities at various chemical potentials remarkably well when confronted with corresponding lattice data. 
Considering that the lattice results have been found by a Taylor expansion in 
powers of the chemical potential, this excellent agreement came as a surprise and indicates rapid
convergence of the power series in $\mu$.

Further developments will be directed towards improvements to overcome some obvious limitations. First, the NJL model operates with a constant four-point coupling strength which supposedly averages the relevant running coupling over a limited low-energy kinematic domain, corresponding to temperatures $T \leq 2 \,T_c$ and chemical potentials $\mu \leq 0.3$ GeV. Contacts with the high-temperature limit of QCD and the HTL approaches need to be established. Secondly, in order to proceed into the range of
larger chemical potentials, diquark degrees of freedom need to be explicitly involved. Also, the effective
potential for the Polyakov loop field, determined so far entirely as a function of temperature by investigating the pure gauge sector, must be examined with respect to its dependence on the chemical potential. And furthermore, the extension to 2+1 flavours with inclusion of strange quarks must be explored.

Nevertheless, considering the simplicity of the PNJL model, the conclusion that can be drawn at this point is promising: it appears that a relatively straightforward quasiparticle approach, with its dynamics rooted in spontaneous chiral symmetry breaking and confinement and with parameters controlled by a few known properties of the gluonic and hadronic sectors of the QCD phase diagram, can account for essential observations from two-flavour $N_c = 3$ Lattice QCD thermodynamics. 
\section*{Acknowledgements}
We gratefully acknowledge stimulating discussions with Jean-Paul Blaizot,
Ulrich Heinz, Volker Koch, Krishna Rajagopal and Helmut Satz. One of us (W.W.) thanks the nuclear theory group at the Lawrence Berkeley National Lab for their kind hospitality. This work was supported in part by INFN and 
BMBF.


\begin{thebibliography}{10}

\bibitem{Fodor:2002km}
Z.~Fodor, S.~D. Katz, and K.~K. Szabo,
\newblock Phys. Lett. B {\bf 568}, 73 (2003).

\bibitem{Fodor:2001pe}
Z.~Fodor and S.~D. Katz,
\newblock JHEP {\bf 0203}, 014 (2002).

\bibitem{Allton:2002zi}
C.~R. Allton {\em et~al.},
\newblock Phys. Rev. D {\bf 66}, 074507 (2002).

\bibitem{Allton:2003vx}
C.~R. Allton {\em et~al.},
\newblock Phys. Rev. D {\bf 68}, 014507 (2003).

\bibitem{Allton:2005gk}
C.~R. Allton {\em et~al.},
\newblock Phys. Rev. D {\bf 71}, 054508 (2005).

\bibitem{Laermann1}
E.~Laermann and O.~Philipsen,
\newblock  Ann.\ Rev.\ Nucl.\ Part.\ Sci.\  {\bf 53}, 163 (2003).

\bibitem{deForcrand:2003hx}
P.~de Forcrand and O.~Philipsen,
Nucl.\ Phys.\ B {\bf 673}, 170 (2003)

\bibitem{delia1}
  M.~D'Elia and M.~P.~Lombardo,
\newblock  Phys.\ Rev.\ D {\bf 67}, 014505 (2003).

\bibitem{delia2}
M.~D'Elia and M.~P.~Lombardo,
\newblock  Phys.\ Rev.\ D {\bf 70}, 074509 (2004).

\bibitem{Arnold:1994eb}
P.~Arnold and C. Zhai,
\newblock Phys. Rev. D {\bf 51}, 1906 (1995).

\bibitem{Zhai:1995ac}
C. Zhai and B. Kastening,
\newblock Phys. Rev. D {\bf 52}, 7232 (1995).

\bibitem{Braaten:1991gm}
E.~Braaten and R.~D. Pisarski,
\newblock Phys. Rev. D {\bf 45}, R1827 (1992).

\bibitem{Frenkel:1991ts}
J.~Frenkel and J.~C. Taylor,
\newblock Nucl. Phys. B {\bf 374}, 156 (1992).

\bibitem{Blaizot:1993be}
J.~P. Blaizot and E.~Iancu,
\newblock Nucl. Phys. B {\bf 417}, 608 (1994).

\bibitem{Andersen:1999fw}
J.~O. Andersen, E.~Braaten, and M.~Strickland,
\newblock Phys. Rev. Lett. {\bf 83}, 2139 (1999).

\bibitem{Andersen:2000zn}
J.~O. Andersen, E.~Braaten, and M.~Strickland,
\newblock Phys. Rev. D {\bf 62}, 045004 (2000).

\bibitem{Andersen:2002ey}
J.~O. Andersen, E.~Braaten, E.~Petitgirard, and M.~Strickland,
\newblock Phys. Rev. D {\bf 66}, 085016 (2002).

\bibitem{Blaizot:1999ip}
J.~P. Blaizot, E.~Iancu, and A.~Rebhan,
\newblock Phys. Rev. Lett. {\bf 83}, 2906 (1999).

\bibitem{Blaizot:1999ap}
J.~P. Blaizot, E.~Iancu, and A.~Rebhan,
\newblock Phys. Lett. B {\bf 470}, 181 (1999).

\bibitem{Blaizot:2000fc}
J.~P. Blaizot, E.~Iancu, and A.~Rebhan,
\newblock Phys. Rev. D {\bf 63}, 065003 (2001).

\bibitem{Kajantie:2002wa}
K.~Kajantie, M.~Laine, K.~Rummukainen, and Y.~Schr\"oder,
\newblock Phys. Rev. D {\bf 67}, 105008 (2003).

\bibitem{Blaizot:2003iq}
J.~P. Blaizot, E.~Iancu, and A.~Rebhan,
\newblock Phys. Rev. D {\bf 68}, 025011 (2003).

\bibitem{Ipp:2003yz}
A.~Ipp, A.~Rebhan, and A.~Vuorinen,
\newblock Phys. Rev. D {\bf 69}, 077901 (2004).

\bibitem{Engels:1982ei}
J.~Engels, F.~Karsch, and H.~Satz,
\newblock Phys. Lett. B {\bf 113}, 398 (1982).

\bibitem{Peshier:1995ty}
A.~Peshier, B.~K\"ampfer, O.~P. Pavlenko, and G.~Soff,
\newblock Phys. Rev. D {\bf 54}, 2399 (1996).

\bibitem{Levai:1997yx}
P.~Levai and U. Heinz,
\newblock Phys. Rev. C {\bf 57}, 1879 (1998).

\bibitem{Peshier:1999ww}
A.~Peshier, B.~K\"ampfer, and G.~Soff,
\newblock Phys. Rev. C {\bf 61}, 045203 (2000).

\bibitem{Szabo:2003kg}
K.~K. Szabo and A.~I. Toth,
\newblock JHEP {\bf 06}, 008 (2003).

\bibitem{Bluhm:2004xn}
M.~Bluhm, B.~K\"ampfer, and G.~Soff,
\newblock J. Phys. G {\bf 31}, S1151 (2005).

\bibitem{Bluhm:2004xn2}
  M.~Bluhm, B.~Kampfer and G.~Soff,
  Phys.\ Lett.\ B {\bf 620} (2005) 131

\bibitem{Pisarski:2000eq}
R.~D. Pisarski,
\newblock Phys. Rev. D {\bf 62}, 111501(R) (2000).

\bibitem{Rebhan:2003wn}
A.~Rebhan and P.~Romatschke,
\newblock Phys. Rev. D {\bf 68}, 025022 (2003).

\bibitem{Schneider:2001nf}
R.~A. Schneider and W.~Weise,
\newblock Phys. Rev. C {\bf 64}, 055201 (2001).

\bibitem{Thaler:2003uz}
M.~A. Thaler, R.~A. Schneider, and W.~Weise,
\newblock Phys. Rev. C {\bf 69}, 035210 (2004).

\bibitem{Drago:2001gd}
A.~Drago, M.~Gibilisco, and C.~Ratti,
\newblock Nucl. Phys. A {\bf 742}, 165 (2004).

\bibitem{Ivanov:2004gq}
Y.~B. Ivanov, V.~V. Skokov, and V.~D. Toneev,
\newblock Phys. Rev. D {\bf 71}, 014005 (2005).

\bibitem{Karsch:2003zq}
F.~Karsch, K.~Redlich and A.~Tawfik,
\newblock Phys. Lett. B {\bf 571}, 67 (2003);
\newblock Eur. Phys. J. C {\bf 29}, 549 (2003).

\bibitem{Rischke:2003mt}
D.~H. Rischke,
\newblock Prog. Part. Nucl. Phys. {\bf 52}, 197 (2004).

\bibitem{Nambu:1961tp}
Y.~Nambu and G.~Jona-Lasinio,
\newblock Phys. Rev. {\bf 122}, 345 (1961).

\bibitem{Nambu:1961fr}
Y.~Nambu and G.~Jona-Lasinio,
\newblock Phys. Rev. {\bf 124}, 246 (1961).

\bibitem{Vogl:1991qt}
U.~Vogl and W.~Weise,
\newblock Prog. Part. Nucl. Phys. {\bf 27}, 195 (1991).

\bibitem{Klevansky:1992qe}
S.~P. Klevansky,
\newblock Rev. Mod. Phys. {\bf 64}, 649 (1992).

\bibitem{Hatsuda:1994pi}
T.~Hatsuda and T.~Kunihiro,
\newblock Phys. Rept. {\bf 247}, 221 (1994).

\bibitem{Buballa}
M.~Buballa,
\newblock Phys. Rept. {\bf 407}, 205 (2005).

\bibitem{Meisinger:1995ih}
P.~N. Meisinger and M.~C. Ogilvie,
\newblock Phys. Lett. B {\bf 379}, 163 (1996).

\bibitem{Meisinger:2001cq}
P.~N. Meisinger, T.~R. Miller, and M.~C. Ogilvie,
\newblock Phys. Rev. D {\bf 65}, 034009 (2002).


\bibitem{Fukushima:2003fw}
K.~Fukushima,
\newblock Phys. Lett. B {\bf 591}, 277 (2004).

\bibitem{Mocsy:2003qw}
  A.~Mocsy, F.~Sannino and K.~Tuominen,
  Phys.\ Rev.\ Lett.\  {\bf 92} (2004) 182302

\bibitem{Megias:2004hj}
  E.~Megias, E.~Ruiz Arriola and L.~L.~Salcedo,
  arXiv:hep-ph/0412308.

\bibitem{Ratti:2004ra}
C.~Ratti and W.~Weise,
\newblock Phys. Rev. D {\bf 70}, 054013 (2004).

\bibitem{Polyakov:1978vu}
A.~M. Polyakov,
\newblock Phys. Lett. B {\bf 72}, 477 (1978).

\bibitem{Susskind:1979up}
L.~Susskind,
\newblock Phys. Rev. D {\bf 20}, 2610 (1979).

\bibitem{Svetitsky:1982gs}
B.~Svetitsky and L.~G. Yaffe,
\newblock Nucl. Phys. B {\bf 210}, 423 (1982).

\bibitem{Svetitsky:1985ye}
B.~Svetitsky,
\newblock Phys. Rept. {\bf 132}, 1 (1986).

\bibitem{Fukushima:2002bk}
K.~Fukushima,
\newblock Ann. Phys. {\bf 304}, 72 (2003).



\bibitem{Dumitru:2005ng}
  A.~Dumitru, R.~D.~Pisarski and D.~Zschiesche,
  Phys.\ Rev.\ D {\bf 72}, 065008 (2005)

\bibitem{Meisinger:2003id}
  P.~N.~Meisinger, M.~C.~Ogilvie and T.~R.~Miller,
  Phys.\ Lett.\ B {\bf 585}, 149 (2004)

\bibitem{Boyd}
G.~Boyd {\em et~al.},
\newblock Nucl. Phys. B {\bf 469}, 419 (1996).


\bibitem{Kaczmarek:2002mc}
O.~Kaczmarek, F.~Karsch, P.~Petreczky, and F.~Zantow,
\newblock Phys. Lett. B {\bf 543}, 41 (2002).

\bibitem{Karsch}
F.~Karsch,
\newblock Lecture Notes in Phys. (Springer) {\bf 583}, 209 (2002);\\
F.~Karsch, F.~Laermann and A.~Peikert,
\newblock Nucl. Phys. B {\bf 605}, 579 (2002).

\bibitem{Fodor}
Z. Fodor, private communication.

\bibitem{Digal}
S.~Digal, F.~Laermann and H.~Satz,
\newblock Eur. Phys. J. C {\bf 18}, 583 (2001).

\bibitem{AliKhan:2001ek}
A.~Ali Khan {\it et al.}  
Phys.\ Rev. D {\bf 64}, 074510 (2001).

\end{thebibliography}
\end{document}